\documentclass[12pt,preprint]{aastex}
\newcommand{\etal} {{\it et al.} }
\newcommand{\bul} {\indent{$\bullet$}}
\newcommand{\arcsecpoint}{\hbox to 1pt{}\rlap{\arcsec}.\hbox to 2pt{}}
\newcommand{\rasecpoint}{\hbox to 1pt{}\rlap{$^s$}.\hbox to 2pt{}}
\newcommand{\arcminpoint}{\hbox to 1pt{}\rlap{\arcmin}.\hbox to 2pt{}}
\begin{document}
\title{VLBI Imaging of Luminous Infrared Galaxies:\\ 
AGN Cores in Mrk231, UGC 5101 \& NGC 7469
\footnote{Based on observations from the VLBA of the National Radio
Astronomy Observatory which is a facility of the US National Science 
Foundation operated under cooperative agreement by Associated Universities,
Inc.}}

\author{CAROL J. LONSDALE}
\affil{Infrared Processing \& Analysis Center\\ 
       California Institute of Technology 100--22\\
       Pasadena, CA 91125;\\
       {\it cjl@ipac.caltech.edu}}

\author{COLIN J. LONSDALE}
\affil{MIT Haystack Observatory, \\
       Westford, MA  01886;\\
       {\it cjl@haystack.mit.edu}}

\author{HARDING E. SMITH }
\affil{Center for Astrophysics \& Space Sciences \\
       University of California, San Diego \\
       La Jolla, CA  92093--0424;\\
       {\it hsmith@ucsd.edu}}

\author{PHILIP J. DIAMOND}
\affil{University of Manchester\\
   Jodrell Bank Observatory    \\
   Macclesfield, Cheshire SK11 9DL, United Kingdom;\\
       {\it pdiamond@jb.man.ac.uk}}
	
	\begin{abstract}
 
We report 18cm VLBI continuum imaging observations at $\sim$5 milliarcsecond
resolution for UGC 5101, NGC 7469, and Mrk 231, all part of a sample of 
Luminous Infrared Galaxies which have been shown to have strong VLBI radio 
cores in a previous 18cm VLBI survey.  Mrk 231, generally considered to be
a dust-enshrouded QSO, shares many characteristics with quasars, including a
broad, Sy1 optical emission spectrum, optical and radio variability, and 
broad-absorption-line (BAL) spectrum.  NGC 7469 is a classical Sy1 galaxy 
and UGC 5101 is a liner system.

The radio morphology of these three systems on VLBI scales is AGN-like, with
well-defined ridgelines and high-brightness yet spatially resolved components.
The structure and flux densities of these VLBI components are not consistent
with starburst generated radio supernovae of the type found in Arp 220.  On
scales of $\sim$100pc, and perhaps beyond, the radio continuum in all three
objects appears to be dominated by an AGN, not a starburst.  Radio emission on
larger scales may well originate, in part or in total, in a less compact
circumnuclear star-forming region.  This is in contrast to the situation
thought to prevail in many LIGs which involves a very compact and dense
nuclear starforming region.  

By placing a lower limit on the AGN-related radio emission, it is possible
to shed light on the relative luminosities of the AGN and starburst in
these objects.  Using the FIR/radio correlation for starburst-related radio
emission, we show that, despite the unambiguous evidence for AGN activity in
these systems, the AGN luminosity, although energetically important, 
may be exceeded by the extended starburst luminosity in all three cases, 
unless the bolometric luminosity of the entire system has been 
underestimated.   The lower 
limits on AGN-related radio emission indicate that these AGNs are 
somewhat over-luminous in the radio compared to radio-quiet PG QSOs, and 
thus that the AGN radio luminosity cannot be used to estimate the bolometric
luminosity of the AGN.  However, comparing near- and mid-IR luminosities of 
the LIGs to PG QSOs we find evidence that a significant fraction of the 
bolometric luminosity of the AGN in Mrk 231, and to lesser extent UGC 5101, 
may be radiated away from Earth and thus not detected at any wavelength.  

The VLBI structure observed in Mrk 231 allows additional interpretation.
Confirming and extending VLBI imaging by \cite{UWC99}, our continuum image
shows a triple structure, with a core and two lobes, classifying it as a
Compact Symmetric Object (CS0).  It has been suggested that these sources
are young, $\tau << 10^6$yr, with the hot spots representing the working
surface of a relativistic jet upon the ambient medium. If the southern
(primary) lobe/hot-spot in Mrk 231 is confined by ram pressure, we estimate
a lobe advance speed, $v_a \sim 10^{-4}c$, and an age for the jet/compact
source, $\tau < 10^6 yr$.

We have also imaged the 1667 MHz OH maser emission in Mrk 231, which is
extended on scales of 50--100 milliarcsec (40--80 pc) and probably coincides
with the inner region of the disk which is seen in CO emission \citep{BS96} 
and HI absorption \citep{CWU98}. Among OH megamaser sources studied at high
sensitivity with milliarcsecond resolution, Mrk 231 is unique in the stringent
upper limits placed upon the flux density of compact OH structures of the type
found in Arp 220 and other LIGs.  It is possible that the circumnuclear
environment of Mrk 231 has been sufficiently disrupted by the emergent QSO 
that the cool, dense clouds necessary for such compact masers no longer exist.

\end{abstract}


	\section{
Introduction}

The most luminous galaxies in the Local Universe are Luminous Infrared
Galaxies (LIGs) which emit the vast majority of their radiant power in the
far-infrared.  These are gas-rich systems which are in the late stages of
collisions or mergers, and extrapolation from the properties of lower
luminosity Starburst galaxies suggests that the LIGs should be active star
forming systems (see Sanders \& Mirabel 1996 for a review).  The LIGs also
show many characteristics of AGN and their luminosities reach values
comparable to those of luminous QSOs.  Much effort has been focused on
whether the power from the LIGs is principally from starburst or AGN
activity, although both types of activity may be present, frequently in the
same system.  The discussion has been framed around a scenario proposed by
\cite{Sand88} in which a merger of gas-rich disk galaxies stimulates a
massive nuclear starburst, which in turn feeds a coalescing AGN core in the
galaxy nucleus.  As the AGN turns on, radiation pressure drives out the
shroud of dust, revealing a nascent quasar.  Other authors \citep{PD85,
SN89, TM85} have proposed variations on the theme of compact, luminous
starbursts giving rise to AGN activity.  The goal must be not only to
understand the dominant source of energy in LIGs, but to understand the
relationship between starburst and AGN activity, and other galaxy
characteristics, and to place them into an evolutionary context.  We have
for some time been approaching this problem by investigating the
characteristics of a complete sample of LIGs defined by Condon \etal\
(1991; CHYT); this work has concentrated on VLBI observations which offer
unique AGN/starburst diagnostics in one of the few wavelength regimes where
the optical depths to the active regions may fall below unity.

In a non-imaging 18-cm VLBI survey of Luminous Infrared Galaxies from the
CHYT sample for compact, high-brightness temperature emission commonly
associated with AGN activity, Lonsdale, Smith, and Lonsdale (1993; Paper I)
showed that milli-arcsecond scale emission with $T_b >> 10^7K$ is common,
perhaps universal in LIGs.  Furthermore, the LIGs follow a similar 
relationship between core radio power and bolometric luminosity to
radio-quiet QSOs \citep{LSL95}.  This work lends support to the
interpretation of LIGs as dust-enshrouded AGN.  On the other hand, in a
detailed analysis of our VLBI survey data, Smith, Lonsdale and Lonsdale
(1998; Paper II) investigated a Starburst origin for LIGs in which the
compact, high-$T_b$ emission is produced by luminous radio supernovae
(RSN).  This analysis indicates that most, but not all, LIG VLBI-scale
emission may be modelled with Starburst-generated RSN, provided the RSN are
{\it extremely luminous}.  The predictions of this analysis were confirmed
with the detection of luminous RSN in the nuclei of Arp 220, consistent
with a Starburst origin for the infrared luminosity of this LIG
\citep{SLLD98}.  We are continuing a monitoring program of the Arp 220 LRSN
allowing refinement of the characteristics of LRSN in the extreme LIG
environment; preliminary results \citep{LDLS01} suggest that decay
timescales are longer than for ``normal'' LRSN with a concomitant reduction
in our estimate of the rate of luminous RSN --- but not total RSN ---
frequency.  Other radio supernova candidates have since been suggested in
galaxies studied here --- 4 near the nucleus of Mrk 231 \citep{T99} and 1
in the starburst ring around the nucleus of NGC 7469 \citep{C01}.

In this paper we present 18cm VLBI imaging of Mrk 231, UGC 5101 and NGC
7469, LIGs which had VLBI detections in our survey of a character which
cannot be adequately fitted by multiple, luminous radio supernovae ---
either isolated or in clumps.  All three galaxies have also commonly been
interpreted as AGN based on optical spectroscopic data. These continuum
imaging observations are designed to determine the nature of the central
radio cores in Mrk 231, UGC 5101, and NGC 7469, to investigate the
relationship between AGN and starburst within these 3 LIGs, and to
investigate the relationship between these galaxies and other AGN and LIG
systems.  We also present 18cm imaging observations of Mrk 231 in the
1667MHz OH maser line.

	\subsection{
UGC 5101}

UGC 5101 (D = 164Mpc; 1${\arcsec}{\sim}$800pc) is a liner system with a 
strongly interacting morphology \citep{Sand88}.  A weak soft Xray detection
by ROSAT has been reported by \cite{RLR96}.  UGC 5101 has an estimated 
reservoir of molecular gas, $log\,M_{H_2} = 10.3 M_\odot$
\citep{Sol97} and a FIR luminosity, $log\,L_{FIR} = 11.96 L_\odot$ 
\citep{Sanders03}.  With a
VLBI flux density estimated from our survey observations, $S_{vlbi} \approx
28$mJy, nearly 20\% of the 18cm flux density was seen in the compact core; 
this is the highest ratio of core-to-total flux density among the 40 LIGs
in our complete sample.

\cite{So00} have presented mid-infrared (MIR) imaging of UGC 5101 using
MIRLIN at Keck and have compared it to IRAS fluxes.  They find a compact
unresolved core of extent less than 0\arcsecpoint 22 ($\sim$175pc) 
at 12.5$\mu$m,
surrounded by more diffuse emission on scales of $\sim 1\arcsec$.  About
60\% of the entire 12$\mu$m flux measured in the large IRAS beam is
contained within in their 4$\arcsec$ measurement, and 30\% of the entire 
12$\mu$m flux is within the unresolved core.  At 17.9$\mu$m the structure is 
consistent with that at 12.5$\mu$m.  

\cite{Sc00} have obtained near-infrared (NIR) HST/NICMOS images of UGC
5101 with resolution about 0\arcsecpoint 2.  The nuclear source in the
near-infrared image is better fit with an r$^{1/4}$ law than an
exponential, with R$_{1/2}<$0.5kpc.  Scoville \etal conclude that although
it is a cool far-infrared source, UGC 5101 has nuclear NIR colors more
similar to those of warm ULIRGs; specifically redder in [$m_{1.6} - m_{2.2}$]
than in [$m_{1.1} - m_{1.6}$] than other cool ULIRGs.  They deduce that a
visual extinction of $A_v >>$ 50 mag is required to explain such colors by a
starburst model in which the gas is mixed with the dust.  The near-infrared
colors are alternatively consistent with a central AGN with high
extinction, or a much less extinguished AGN plus warm dust emission at
2.2$\mu$m.

\cite{So00} compare their 12$\mu$m image with the 
NICMOS NIR image of \cite{Sc00} and the VLA 8.4GHz image of
CHYT, finding the sources aligned at these three frequencies, 
insofar as the relative astrometry allows.  

\cite{T01} have analysed the 3\arcmin aperture ISO CVF observations
of the MIR spectrum of UGC 5101 in terms of a starburst and an AGN
contribution, based on the ratio of the strength of the 7.7$\mu$m
unidentified feature relative to the continuum, combined with the
5--10$\mu$m continuum colors.  They conclude that approximately 50\% of the
MIR continuum could come from an AGN, which is consistent with the
result of Soifer \etal that $\sim 30$\% of the the total IRAS 12$\mu$m flux
comes from the unresolved ($<$ 0\arcsecpoint 22) MIR core.

\cite{Spoon02} report detection of the
6.0$\mu$m water ice feature in ISO-PHT-S spectroscopy of UGC 5101, the
strongest amongst the 20 galaxies in their sample, as well as strong
9.7$\mu$m silicate absorption and PAH emission features.  Within the large
ISO beam it is not possible to determine the scale on which the absorption
is occurring.  They caution that the detailed structure of the absorption
features and MIR continuum complicates the attempt to separate the
starburst vs AGN contributions to the luminosity by Tran \etal.  Also, the
presence of the deep silicate feature may affect Soifer {\it et al.}'s
conclusion that only 30\% of the total IRAS flux arises from the unresolved
core.  Moreover, both the Tran \etal and Soifer \etal studies could have
underestimated the total MIR radiation arising from the AGN if the
extinction to the nucleus is significant in the MIR.

	\subsection{
NGC 7469}

NGC 7469 (D = 66 Mpc) is a classical Seyfert Galaxy \citep{Sy43} which has
subsequently been classified as a Sy1--1.2 galaxy based on the strengths of
the broad and narrow components of its optical emission lines.  NGC
7469 is strongly interacting with a nearby companion, IC5283 \citep{D85}.
The galaxy shows a strongly variable UV/optical continuum with low extinction,
and X-ray emission which is variable on time-scales of $\sim$1 day 
\citep{N00} with a complex iron line spectrum from BeppoSAX \citep{DeRosa02}.
There is evidence for an $\sim$ 10$^7$ M$_{\odot}$
central black hole from reverberation mapping \citep{PW00}.  NGC 7469 shows
nuclear non-thermal radio
emission as expected for classical AGN, but also shows a dual ring of star
formation 1--3{\arcsec} and 8--10{\arcsec} from the nucleus.  The inner
starburst ring is detected at radio \citep{W91}, mid-infrared
\citep{Miles94, lws, So02} and optical wavelengths \citep{M94}.  

About half of the 12\micron\ emission originates in this starburst ring, 
which also shows the 11.3\micron\ emission feature commonly associated with
regions of strong star formation, but the ratio of nuclear ring emission
increases at longer wavelengths, suggesting that the nucleus may dominate
in the FIR \citep{lws}.  NICMOS observations of NGC 7469 by Scoville \etal 
reveal that $>$30\% of the 2.2$\mu$m flux from the galaxy arises within a
radius of 0$\arcsecpoint$12, or $\la$77pc.  \cite{Alon01} determine
non-stellar nuclear fluxes for NGC 7469 between 1--5$\mu$m, which together
with ISO fluxes they model in terms of a close to face-on dusty molecular
torus.

The system is gas-rich with $M_{H_2} = 10^{10} M_\odot$
\citep{Sand91}.  Our VLBI survey observations \citep{LSL93} 
showed a strong compact radio core with an estimated flux density of 12mJy,
about 6\% of the integrated 18cm radio emission, and an estimated
brightness temperature in excess of $10^7$K.  MERLIN observations at 18cm
by \cite{TGPK01} show a barely resolved nuclear source with
$\sim$30 mJy, surrounded by more diffuse emission associated with the inner
starburst ring.  This result implies that our VLBI survey observations were
insensitive to 15--20 mJy of emission on scales $\la$0.1 arcsec.  A
careful registration of the new 11.7$\mu$m image of NGC 7469 of \cite{So02}
with the archived MERLIN 18cm map of \cite{TGPK01} indicates that the
compact radio and mid-infrared cores align to better than $\sim$80mas; see
\cite{So02}.

	\subsection{ 
Mrk 231}

Mrk 231 (=UGC 5058; $D = 173$ Mpc; 1\arcsec = 840pc for $H_0 =
75$km\,s$^{-1}$Mpc$^{-1}$) has long been recognized as a remarkable galaxy:
it is the most luminous galaxy in the local ($z \la 0.1$) Universe with
characteristics that place it among classical AGN \citep{B77} and also
infrared galaxies \citep{Sand88}.  Our interest in Mrk 231 was stimulated
by the 1993 18cm VLBI survey \citep{LSL93} which demonstrated that Mrk 231
has a strong, high-brightness temperature, VLBI core radio source which
accounts for a major fraction of its 18cm radio emission. In its radio
properties Mrk 231 is thus the most ``quasar-like'' of our VLBI sample and
of the complete sample of Luminous Infrared Galaxies \citep{CHYT}
from which our sample was derived.  

\subsubsection{Mrk 231 as a QSO}

Mrk 231 has long been considered an infrared quasar --- the integrated
luminosity of Mrk 231, $L_{bol} \ga 10^{46} erg\,s^{-1}$, is comparable
to other low-redshift ($z \la 1 $) radio-quiet QSOs and it falls in the
midst of the radio power-luminosity relation for QSOs (and LIGs)
constructed by \cite{LSL95}.  Mrk 231 is reported to be variable at optical 
\citep{HK87}, radio \citep{MG78}, and X-ray wavelengths \citep{Gal02}.

The core of Mrk 231 has been imaged by HST  at 1.6$\mu$m \citep{Q01}, 
revealing an unresolved point source, FWHM$\la$0.13{\arcsec} (109pc) with 
absolute magnitude $M_H \sim -26$, which is comparable to the most luminous 
z$<$0.2 PG QSOs.  Adaptive optics imaging by \cite{Lai98} detected a point 
source of FWHM 0.1{\arcsec}, with $[J-H]=0.99$, $[H-K]=1.12$.   
Mrk 231 shows a strong, broad optical type 1 emission-line spectrum with 
strong FeII emission.  The optical lines and continuum are reddened 
with $A_V \approx 2$ \citep{B77, LCM94}, and the optical/NIR continuum
polarization level decreases from as high as 20\% in the UV towards longer 
wavelengths \citep{S95}, interpreted as a complex scattering geometry in 
which most lines of sight to the nucleus are moderately obscured.

Mrk 231 was imaged at Keck in 7 MIR bands from 7.9 to 19.7$\mu$m by 
\cite{So00},
and found to be unresolved at 0.13{\arcsec} in all bands; {\it i.e.} on a 
similar 100pc scale as the NIR point source.   All of the 12$\mu$m IRAS
emission in a much bigger beam is contained within this unresolved source.
\cite{Spoon02} have recently
published an ISO-PHT-S MIR spectrum of Mrk 231, which shows strong 
silicate 9.7$\mu$m absorption, and probably water ice 6.0$\mu$m absorption,
interpreted as a molecular torus around the AGN.  The extinction to the
AGN cannot be directly estimated from the depth of the silicate absorption
because it depends strongly upon viewing angle, involving radiative transfer
through the few pc-diameter torus \citep{ERR95}.

Mrk 231 is a low-ionization BAL (Broad Absorption Line) object with three
systems: $v_{ej} \approx 4700\,km/s$, $v_{ej} \approx 6000\,km/s$, $v_{ej}
\approx 8000\,km/s$ \citep{AW72, B77, RFS85}. The $v_{ej} \approx 8000\,km/s$ 
absorption-line system is variable on timescales of order 2--3 yrs
\citep{B91, FRM95}.  Mrk 231 is underluminous in x-rays between 0.1--10kev
\citep{MR00, Gal02} and soft $\gamma-$rays \citep{D97} a
characteristic it shares with other BAL objects \citep{Gal99}.
Spectroscopy with Chandra \citep{Gal02} in the 2--10kev range reveals a hard
spectrum, interpreted as several scattered emission components, and rapid
variability which ties the emission to within light-hours of the central
black hole.  Neither Chandra nor ASCA \citep{MR00} directly detects
a hard Xray power-law, indicating the presence of a Compton-thick
absorber, $N_H > 10^{24}$.
\cite{Gal02} suggest the absorber could be the so-called ``hitchhiking'' gas 
at very small radii ($\sim 10^{15}$cm), invoked to protect the BAL wind 
from ionization by the UV and soft-X-ray photons \citep{Mur95}.  The 
X-ray absorber must be small to avoid blocking the highly 
variable light-hour-scale scattered X-ray component from view.

\subsubsection{Mrk 231 as a Luminous Infrared Galaxy}

With an infrared luminosity, $log\,L_{FIR} = 12.51 L_\odot$ 
\citep{Sanders03}, Mrk 231 is
the most luminous infrared galaxy in the complete sample of 40 LIGs of
CHYT from the IRAS Bright Galaxy Sample.  Like Arp
220 and other well-studied LIGs, the system is rich in molecular gas,
$log\,M_{H_2}\, \approx 10.2 M_\odot$ \citep{SYB97} and it is one of
the original OH megamaser systems \citep{Baan85}.  Furthermore, Mrk 231
shows tidal tails and other evidence for merger or disturbance in the
``host'' galaxy, which is itself a luminous system with evidence for rapid
star formation \citep{HK87, Sand88}.

The CO in Mrk 231 has been studied at high resolution by \cite{BS96}
who show that the CO distribution is elongated with a velocity
gradient aligned with the major axis of the CO distribution.  They
interpret the CO distribution as a disk, oriented roughly E--W, containing
an estimated $3\times 10^9 M_{\odot}$ of molecular gas within a diameter of
about 1\arcsec(=840pc) and a scale height, $h \approx 30pc$.  Bryant \&
Scoville find the CO brightness temperature to be comparable to the FIR
dust temperature, $T_b(CO) \sim T_{fir} \sim 85K$ and suggest that the disk
is filled with dense gas clouds, which may account for the accretion rate
required to power the luminous nucleus.

In a comprehensive series of papers Ulvestad and colleagues have described
results from a multi-frequency set of VLBA/VLA observations of Mrk 231.
\cite{UWC99} present observations from 1.4--15GHz
which show a central north-south triple source with overall extent $\la
100$mas (80pc), indicative of an AGN in structure.  
Both lobes of the north-south triple show evidence for
free-free absorption near 2 GHz which is interpreted as due to ionized gas 
with $n_e \sim 10^3$cm$^{-3}$.  The central (nuclear) component of the
triple shows resolution above 5GHz and
is resolved into 2 or possibly 3 components at 15GHz with extent $\sim
3$mas (2pc) at PA $65^\circ$.

21cm line continuum observations with the VLBA \citep{CWU98}
show emission from an extended ``disk'' with a size of 440mas (370pc) which
exhibits HI {\it absorption} against radio emission from the inner disk 
($r \la 160pc$), but not against the central radio core.  
The HI observations suggest $N_{HI} \approx 6\times 10^{22}
 (T_s/1000K) cm^{-2}$; $<n_{HI}> \approx 250 (T_s/1000K)\,cm^{-3}$; and
$M_{HI} \approx 2 \times 10^7 (T_s/1000) M_\odot$, where $T_s$ is the spin
temperature.  These observations are
consistent with the neutral hydrogen lying in a disk structure of sub-kpc
dimension.  Further continuum observations reported by \cite{T99}
trace this emission out to nearly 1''.  They interpret the disk radio 
emission as due to star-formation activity, $\dot m \sim
60-220\,M_\odot\,yr^{-1}$.  

On larger scales \cite{UWC99} find continuum emission emerging
from the core near star-forming knots \citep{HST}, extending to the south 
for about 20\arcsec, coincident with regions of H$\alpha$ emission
\citep{HK87}, then curving toward the west in the general detection of
the tidal tails observed by \cite{HK87} and \cite{HST}.  They interpret 
this kpc-scale radio
emission as nonthermal emission powered by a 
jet from the nucleus, principally due to the high degree of linear 
polarization and the highly ordered magnetic field implied.  
This emission has previously been interpreted as due to a starburst-driven
superwind by \cite{B93}.

	\section{
VLBI Observations}

Observations of UGC 5101 and NGC 7469 were obtained as part of VLBA
continuum imaging experiment BL6 on 16 January 1995.  All 10 telescopes
of the VLBA were used, with on-source integration times of roughly
4 hours per source.  UGC 5101 and, in particular, NGC 7469 are relatively
weak sources on VLBI scales, as are most of the LIGs in our original
VLBI Survey sample, so conventional self-calibration and hybrid mapping 
techniques were not readily applicable.  

However, unrelated compact radio sources bright enough for VLBI 
fringe-fitting lie within a few arcminutes of both galaxies, and it was 
possible to perform in-beam phase referencing to calibrate the phases of 
the target sources.  In fact,
UGC 5101 and NGC 7469 were selected for this imaging program as having
nearby calibrators within the 30\arcmin primary beam from a survey of our 
original list of LIGs with compact VLBI-scale radio emission.  The phase 
solutions from SELFCAL show a point-to-point scatter smaller than 10$^\circ$,
commensurate with the signal-to-noise ratio of about 5 for these solutions,
and these solutions appear adequate for well over 95\% of the data.
Long-term trends in phase consistent with typical L-band
atmospheric/ionospheric drifts are evident and benign.
In this
way we were able to produce accurate maps with peak flux densities as low as
1.4 mJy.  Typical background rms levels of 60$\mu$Jy for NGC 7469 and 
80$\mu$Jy for UGC 5101 were achieved.

\clearpage

\begin{deluxetable}{l l r r c c r}
\tabletypesize \footnotesize
\tablenum{1}
\tablewidth{0pt} 
\tablecaption{Calibrators for VLBI Experiment BL6}
\tablehead{
\colhead{VLBI Source} &  
	\colhead{Calibrator} & \colhead{$\alpha$} & \colhead{$\delta$} & 
	 \colhead{RA Offset} &\colhead{Dec Offset} &\colhead{$S_{1.67}^{pk}$}\\
	\colhead{} & \colhead{} & \colhead{} & \colhead{} & 
         \colhead{} & \colhead{} & \colhead{(mJy)} \\
\colhead{(1)} & \colhead{(2)} & \colhead{(3)} & \colhead{(4)} & 
	\colhead{(5)}  & \colhead{(6)}  & \colhead{(7)}\\
}
\startdata
UGC 5101 & 0935+6117 & $09^h\ 35^m$ 30\rasecpoint 865 & 
           $+61^\circ$ 17\arcmin\ 24\arcsecpoint 129 & 2\arcmin\ 31\arcsec &
           3\arcmin\ 48\arcsec & 34 \\
NGC 7469 & 2303+0850 & $23^h\ 03^m$ 11\rasecpoint 912 & 
           $+08^\circ$ 50\arcmin\ 13\arcsecpoint 029 & 0\arcmin\ 54\arcsec &
           2\arcmin\ 13\arcsec & 7 \\
\enddata
\end{deluxetable}

\clearpage

Emission was detected on a variety of size scales, particularly in UGC 5101.
The surface brightness sensitivity of the images is a strong function of
the data weighting during the mapping process, so we present two
images of UGC 5101, one optimizing angular resolution for the purpose of
estimating the sizes of compact features, and one with lower resolution, but a
better representation of diffuse emission.

The observations of Mrk 231, conducted under project code GL15 on 13 November
1994, involving 17 telescopes in Europe and the U.S., were designed to provide
full imaging in line and continuum for the four brightest OH megamaser sources
in the northern sky, including IIIZw35, Mrk 231, Arp 220 and IRAS17208$-$0014.
Mrk 231 received a long track, of which $\sim$4 hours was spent on-source,
yielding excellent u--v coverage.  After correlation in April 1996, standard
hybrid mapping techniques were used to reduce the observations of this
relatively bright source.  The rms noise levels achieved were close to
60$\mu$Jy/beam.  As for UGC 5101, we show two images, one optimized for angular
resolution, and the other to better delineate more extended emission.

Our VLBI maps of the three galaxies are shown in Figures 1--3: Figure 1
shows UGC 5101; (1a) is a near-uniform weighting map exhibiting the highest
resolution (with gaussian restoring beam 7.6$\times$6.2mas in PA 24$^\circ$
and (1b) is a map tapered to display extended emission (convolution size
11.6$\times$9.9mas in PA 42$^\circ$).  Figure 2 shows the map of NGC 7469
(convolution size 13.7$\times$7.9mas in PA $-8^\circ$).  Figure 3 shows Mrk
231; (3a) is the uniformly weighted image with a 4.8$\times$3.4 milliarcsec
beam, and (3b) is the naturally weighted image with a 6.1$\times$4.2
milliarcsec beam.

	\section{
Results}

	\subsection{
UGC 5101}

The image of UGC 5101 (Figure 1) displays three major compact, high
brightness temperature components in the NW, and a more diffuse component
in the SE.  The compact components are barely resolved,
implying diameters $\la 3$ -- 4pc.  There is faint diffuse emission
connecting the 3 compact components, which lie within an area of $\sim 60
\times 30$ mas (RA $\times$ Dec), or 48 $\times$ 24 pc.  The SE diffuse
component, A, lies about 60 pc from the easternmost compact component, and a
wisp of diffuse emission extends from it towards the compact components.
Component A is clearly associated with a spur of emission visible on the
highest resolution VLA images of SLL98 on 0{\arcsecpoint}1 scales.   
It is likely
that the flux density of this diffuse structure has been significantly 
underestimated by our VLBI observations.
Details for the compact components are given in Table 2: column (1) gives
our component designation; columns (2) and (3) the positional offset
from component ``C''; column (4)
the peak 1.67GHz flux density; column (5) the integrated
1.67GHz flux density; columns (6), (7), (8) the major and minor axial
extent and the position angle of an elliptical gaussian fit from AIPS
procedure JMFIT and column (9) the integrated radio luminosity
(10MHz--100GHz) from standard synchrotron theory \citep{synch}.

The major components lie along a discernible ridge line; at the three compact 
components the ridge line is at approximate position angle 107$^{\circ}$, then 
it bends to $\sim 135^{\circ}$ towards the diffuse component.  The total extent
of the VLBI-scale structure is 0{\arcsecpoint}13 $\times$ 0{\arcsecpoint}12 
(RA $\times$ Dec), which corresponds to 103 x 95 pc at the distance of 164 
Mpc adopted for UGC 5101.   Our experiment is
not sensitive to diffuse extended structures on scales significantly 
larger than these.  The total flux detected in our experiment is 38 mJy,
which is over 25\% of the 18 cm emission detected on scales of 1{\arcsecpoint}5
with the VLA \citep{CHYT}.

The fact that the major components form a rough ridge-line, coupled with the
evidence for diffuse emission connecting the components, strongly suggests 
an AGN-generated jet structure.  If this is the correct interpretation
for these features, one of the compact sources is likely to be the core and
the others to be knots in the jet.  We cannot tell from our single frequency 
data which of the compact components is likely to be the core.  The diffuse
component may be due to strong interaction between the jet and surrounding
ISM, decollimating and bending the flow.  There is no evidence for any bow
shock or limb brightnening in the diffuse component.  The appearance is
reminiscent of centrally-peaked Fanaroff and Riley type I morphology
associated with low-luminosity extragalactic radio sources.

The general appearance of the UGC 5101 VLBI structures resembles 
compact jets in other AGN observed with 18cm VLBI --- {\it e.g.}
0402+379=4C37.11, 0831+557=4C55.16, 0900+428=4C42.28, {\it etc.}, in the
Caltech Jodrell Bank VLBI Survey \citep{P95}.

It is very unlikely that the compact structures in UGC 5101 are radio
supernovae, (RSN) such as those we discovered in Arp 220 \citep{SLLD98}.
The individual UGC 5101 components are up to $\sim$30 times more luminous
than the brightest Arp 220 RSN.  Moreover evidence for extension of the
compact components indicates source sizes much too large for a single RSN,
which expand at a rate of about 0.01 pc$\,$yr$^{-1}$
(0.013 mas$\,$yr$^{-1}$) near maximum, corresponding to minimum ages
exceeding 300yr for resolved features.  Neither can clusters of RSN
straightforwardly explain the compact emission: the RSN rate implied by the
FIR luminosity of UGC 5101 is $\nu_{sn} \approx 1\,$yr$^{-1}$ comparable to
that for Arp220 \citep{SLL98}, but the components A--D are each factors of
several greater in radio power than the total 1.67GHz radio power of the
Arp220 RSN \citep{SLLD98}.

\clearpage

\begin{figure}
\figurenum{1}
\plottwo{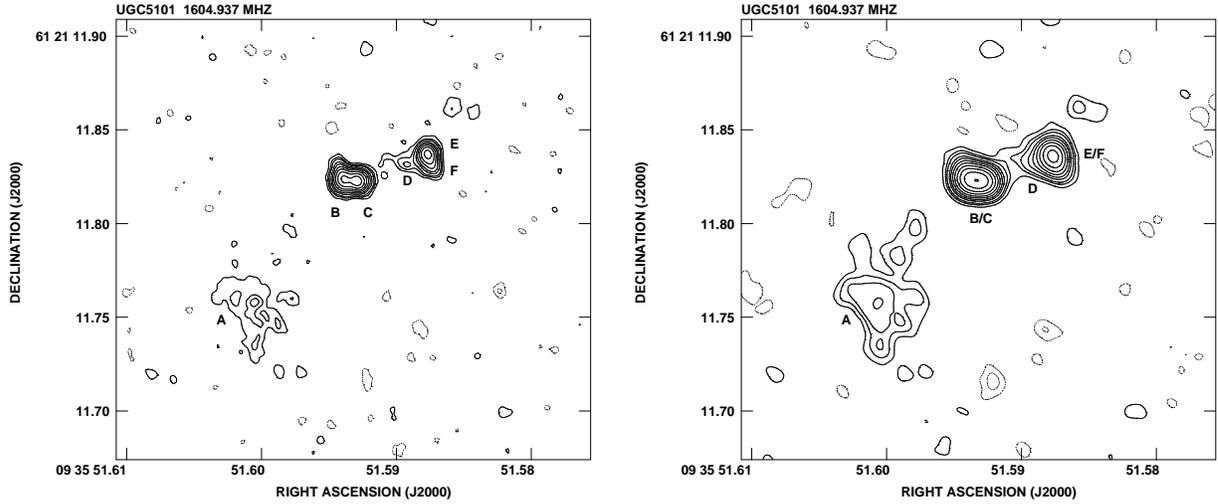}{f1b.eps}
\caption{
VLBI continuum images of UGC 5101.  (1a) is a near-uniform weighting map 
exhibiting the highest resolution (with gaussian restoring beam 
7.6$\times$6.2mas in PA 24$^\circ$; (1b) is a map tapered to display 
extended emission (convolution size 11.6$\times$9.9mas in PA 42$^\circ$).
Contours are -0.6, -0.4, -0.2, 0.2, 0.4, 0.6, 1.0, 1.4, 2.0, 2.8, 4.0, 5.6,
8.0, 11.2mJy.}
\end{figure}

\clearpage

\begin{deluxetable}{r r r r r r r r r r}
\tabletypesize \footnotesize
\tablenum{2}
\tablewidth{0pt} 
\tablecaption{COMPACT 18cm COMPONENTS IN UGC 5101}
\tablehead{
\colhead{Component} &  
	\colhead{$\Delta\alpha$} & \colhead{$\Delta\delta$} & 
	\colhead{$S_{1.67}^{pk}$} & \colhead{$S_{1.67}^{int}$} & 
	\colhead{$\theta_{maj}$} & \colhead{$\theta_{min}$} & 
	\colhead{p.a.} & \colhead{$L_{radio}$}  \\
	\colhead{} & \colhead{(mas)} & \colhead{(mas)} & \colhead{(mJy)} & 
	\colhead{mJy} & \colhead{(mas)} & \colhead{(mas)} & 
	\colhead{($^\circ$)} & \colhead{($erg\,s^{-1}$)}  \\
\colhead{(1)} & \colhead{(2)} & \colhead{(3)} & \colhead{(4)} & 
	\colhead{(5)}  & \colhead{(6)}  & \colhead{(7)}  & \colhead{(8)}  & 
	\colhead{(9)}  \\
}
\startdata
&&&&\multicolumn{2}{c}{$S_{vlbi} = 38$mJy}&&&\\
 A & $\sim$55 & $\sim -65$ & \nodata     & $9.6\pm0.7$ & \nodata & \nodata  
	& \nodata  & $5.4\times 10^{39}$  \\
 B &    8.1   &  1.2      & $5.2\pm0.1$ & $8.2\pm0.5$ & $< 8$ & $< 8$  
	& 177:     & $4.7\times 10^{39}$  \\
 C &    0    &   0     & $6.8\pm0.1$ & $10.4\pm0.5$ & $6\pm1$ & $< 5$  
	& $104\pm40$ & $6.2\times 10^{39}$  \\
 D &  $\sim$25 &  $\sim$10 & \nodata    & $0.8\pm0.1$ & \nodata & \nodata 
	& \nodata  & \nodata              \\
 E &   37.7  & 15.2     & $3.8\pm0.1$ & $4.5\pm1.0$ & $< 6$        & $< 6$
	& \nodata  & $2.6\times 10^{39}$  \\
 F &   38.3  &  9.2     & $2.2\pm0.1$ & $4.5\pm1.0$ & $< 10$ & $< 10$
	& $54\pm5$ & $2.4\times 10^{39}$  \\
\enddata
\end{deluxetable}

\clearpage

The upper limit to the size of the mid-infrared core source detected by 
\cite{So00} --- $\la$175pc --- is comparable to the total size of our 
18 cm source, and is coincident with it within the positional uncertainty
of the mid-infrared image.

	\subsection{
NGC 7469}

The image of NGC 7469 displays 5 compact sources which lie roughly on
an E--W line.  The 2 western sources are well resolved, and the westernmost
could itself be double.  Two of the 3 eastern sources are unresolved,
and the central of these three has marginal evidence for resolution.   
There is one additional possible source to the south at low SNR, with flux
about 0.3 mJy.   See Table 3 for details; column legend is as for Table 2. 

There is no VLBI evidence for diffuse emission, connecting ridges or jets
between the compact sources.  The source complex is contained within an area 
of dimension $\sim$168 mas ($\sim$54 pc) EW.  Diffuse emission on 
scales greater than about 0{\arcsecpoint}1 ($\sim$30 pc) would not 
have been detected by our
experiment.  However, the total flux density detected in our experiment is
about 15 mJy, which is $\sim$50\% of the flux density seen by \cite{TGPK01}
within a $\sim$0{\arcsecpoint}15 nuclear component at 18cm using MERLIN.
Their contour map indicates clear evidence for E--W extension of the nuclear
component with a size and orientation consistent with our VLBI-scale structure.
The MERLIN data indicate that our VLBI components are embedded in a region of
somewhat more diffuse emission with a scale of order 150 mas (48 pc), 
which was
resolved out on the VLBI baselines.  The small size of the MERLIN component
implies that the compact and diffuse components of the structure are coincident
on the sky ({\it i.e.}  the compact VLBI components are embedded within the diffuse
emission).  Furthermore, the ``diffuse" MERLIN component is itself compact by
the standards of starburst-related diffuse emission, with a brightness
temperature approaching 10$^6$K, and is thus not readily associated with
star-formation as opposed to an AGN-related radio source.

The unresolved sources could in principle be individual RSN like
those in Arp 220 \citep{SLLD98}, being up to a factor of two more
luminous than the Arp 220 RSN; the resolved western features, which are
too large to be individual RSN (see Table 2), could be clumps of luminous
RSN.  We consider this unlikely for several reasons: 1) the inferred
supernova rate in NGC 7469,  ascribing {\it all} infrared emission to a
Starburst origin, is about a factor of 5 lower than for Arp 220
\citep{SLL98}, 2) reasonable combinations of supernova luminosity and
age would require some resolution of such a clump of RSN, and  3)
one would expect compact AGN emission coincident with the Sy 1 nucleus
whereas RSN would be expected to be found in the Starburst
ring.  The most plausible hypothesis, which we adopt, is that the
small-scale nuclear radio emission in NGC 7469 is AGN-related.

\clearpage

\begin{figure}
\figurenum{2}
\epsscale{0.60}
\plotone{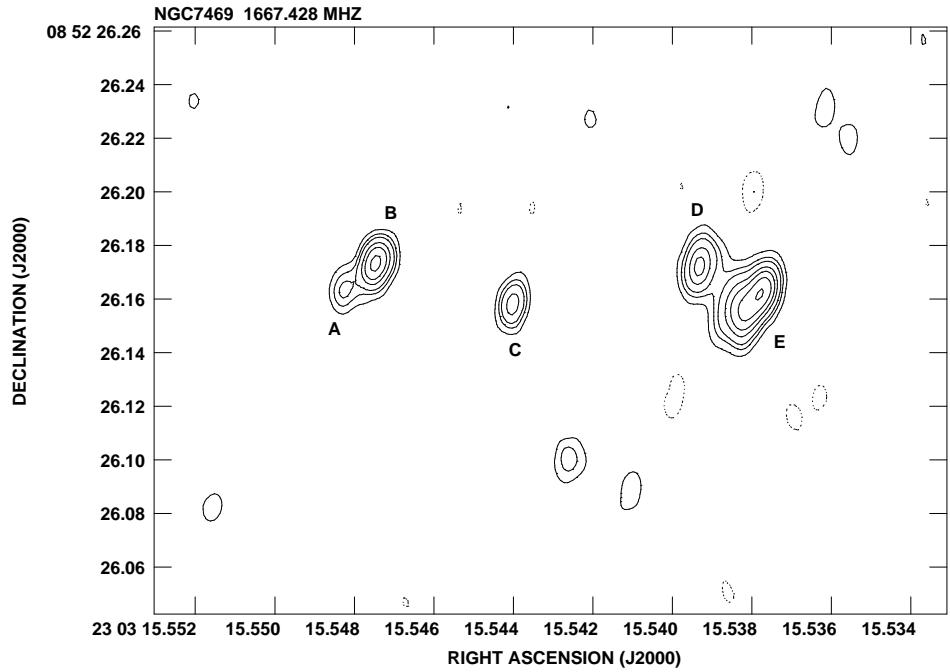}
\caption{
1.67MHz near-uniform weighting map of NGC 7469 (convolution size 
13.7$\times$7.9mas in PA $-8^\circ$).  Tapering reveals no further extended
emission. Contours are -0.3, -0.2, -0.1, 0.1, 0.2, 0.3, 0.5, 0.7, 1.0, 1.4mJy.}
\end{figure}

\clearpage

\begin{deluxetable}{r r r r r r r r r}
\tabletypesize \footnotesize
\tablenum{3}
\tablewidth{0pt} 
\tablecaption{COMPACT 18cm COMPONENTS IN NGC 7469}
\tablehead{
\colhead{Component} &  
	\colhead{$\Delta\alpha$} & \colhead{$\Delta\delta$} & 
	\colhead{$S_{1.67}^{pk}$} & \colhead{$S_{1.67}^{int}$} & 
	\colhead{$\theta_{maj}$} & \colhead{$\theta_{min}$} & 
	\colhead{p.a.} & \colhead{$L_{radio}$} \\
	\colhead{} & \colhead{(mas)} & \colhead{(mas)} & \colhead{(mJy)} & 
	\colhead{mJy} & \colhead{(mas)} & \colhead{(mas)} & 
	\colhead{($^\circ$)} & \colhead{($erg\,s^{-1}$)} \\
\colhead{(1)} & \colhead{(2)} & \colhead{(3)} & \colhead{(4)} & 
	\colhead{(5)}  & \colhead{(6)}  & \colhead{(7)}  & \colhead{(8)}  & 
	\colhead{(9)}
}
\startdata
&&&&\multicolumn{2}{c}{$S_{vlbi} = 14.8$mJy}&&&\\
 A &  54.2 &  6.3  & $0.7\pm0.1$ & $1.4\pm0.2$ & $<19$    & $<19$
	& \nodata   & $1.3\times 10^{38}$ \\
 B &  51.2 & 15.2  & $2.1\pm0.1$ & $2.7\pm0.2$ & $<8$     & $<8$  
	& \nodata   & $2.5\times 10^{38}$ \\
 C &   0   &  0    & $1.2\pm0.1$ & $1.2\pm0.1$ & $<5$     & $<5$  
	& \nodata   & $1.1\times 10^{38}$ \\
 D & $-$69.8 & 14.1  & $1.3\pm0.1$ & $2.8\pm0.2$ & $13\pm1$ & $8\pm1$ 
	& $16\pm6$  & $2.5\times 10^{38}$ \\
 E & $-$89.7 &  1.5  & $2.6\pm0.1$ & $7.4\pm0.3$ & $21\pm1$ & $7\pm1$
	& $135\pm5$  & $6.9\times 10^{38}$ \\
\enddata
\end{deluxetable}

\clearpage

The Scoville \etal NICMOS images of NGC 7469 have a limit on the core size
$\la$77pc which is comparable to the overall dimension of the 
structures detected in our images: $\sim$54pc.   The 12$\mu$m core is 
similarly self-consistent in size to the VLBI image, and is coincident
with it to $\sim$40mas (13pc) \citep{So02}.

	\subsection{Mrk 231}

Our global VLBI experiment exhibits better resolution and somewhat lower 
noise than the 1.4GHz VLBA image of \cite{UWC99}, but generally
confirms their results.  The central 100mas exhibits a compact core with
roughly symmetrically placed lobes approximately 30mas north and south.
The southern lobe exhibits strong brightening at the extreme limb, which
almost certainly represents the bow shock at the working surface of a
nuclear jet upon the ambient medium.

	\subsubsection{
The Compact AGN Continuum Source in Mrk 231}

The radio morphology of the central region of Mrk 231 ($\theta \la
0\arcsecpoint 1$, $d \la 84$pc) places it among the ``Compact Symmetric
Objects'' --- CSOs \citep{W94} --- in which radio emission originates from
hot spots or lobes symmetrically placed on sub-kpc scales about a central
compact source.  \cite{R96} list a dozen characteristics which are
exhibited by all or some of the 5 confirmed CSOs in the \cite{PR88} and
Caltech-Jodrell Bank \citep{P95, H95} VLBI samples, including
high-luminosity, variability, optical evidence for interaction (in one
case), etc.  Mrk 231 shares most of these characteristics with the
exception that in Mrk 231 the core fraction is much higher than in the
confirmed CSOs and that other CSOs appear to be in E galaxies dominated by
an old stellar population, whereas Mrk 231 is a gas-rich system with a
strong starburst.  Also, by the above definition a common characteristic 
of CSOs is lack of extended emission on scales larger than 1 kpc, whereas 
Mrk 231 
shows extended emission due perhaps to a superwind from the circumnuclear
starburst \citep{UWC99}.

It has
been suggested that the CSOs are young, $\tau << 10^6$yr, with the hot
spots representing the working surface of a relativistic jet upon the
ambient medium. In this picture the hot spots are confined by ram pressure,
advancing with inferred typical speed, $v_a \sim 0.02c$ \citep{R96};
advance speeds measured by \cite{T00} for 4 CSOs are an order of magnitude
larger, of the order of 0.1--0.5$h^{-1}\,c$, with kinematic ages ranging
from 300 -- 1200 yr.

\clearpage

\begin{figure}
\figurenum{3}
\plottwo{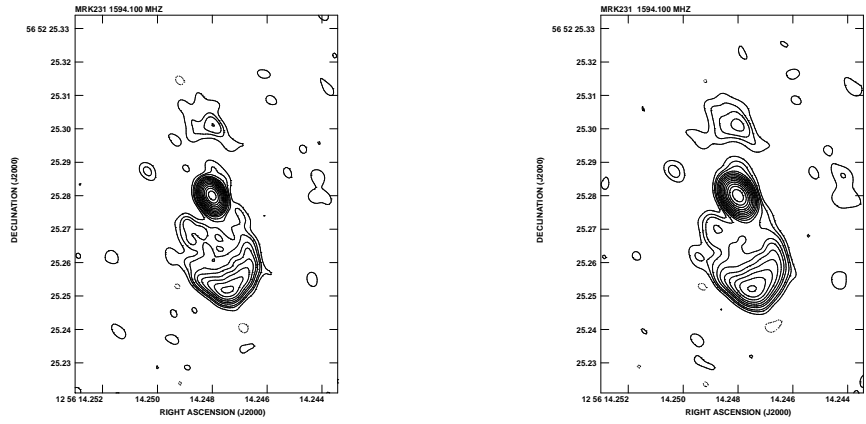}{f3b.eps}
\caption{
VLBI continuum images of Mrk 231. (3a) is the uniformly weighted image 
with a 4.8$\times$3.4 milliarcsec
beam; (3b) is the naturally weighted image with a 6.1$\times$4.2
milliarcsec beam.  Contours are -0.6, -0.4, -0.2, 0.2, 0.4, 0.6, 1.0, 1.4, 
2.0, 2.8, 4.0, 5.6, 8.0, 11.2, 16.0, 22.4, 32.0mJy}

\end{figure}

\clearpage

\begin{deluxetable}{l c c r l}
\tablenum{4}
\tablewidth{0pt} 
\tablecaption{Mrk 231 RADIO COMPONENTS}
\tablehead{
\colhead{Component} & \colhead{Scale} & \omit         & \colhead{$S_{1.6GHz}$} 
& \colhead{Reference}\\
\omit	            &  \omit          & \omit         & \colhead{(mJy)} &
\omit \\
\colhead{(1)}       &  \colhead{(2)}  & \colhead{(3)} & \colhead{(4)}   & 
\colhead{(5)}\\}
\startdata
 Extended   & $> 1\arcsec$ & Superwind? & 42 & Ulvestad,
Wrobel \& Carilli (1999)\\
            & 0\arcsecpoint 1 -- 1\arcsecpoint 0 & Starburst/Molecular Disk &
130 & Carilli, Wrobel \& Ulvestad (1998)\\
 Core       & $< 0\arcsecpoint 005$ & AGN Core & 43$\pm 4$ & This work\\
 South Lobe &                       & Jet Interaction Region & 44$\pm 6$ & This work\\
 North Lobe &                       & Jet Interaction Region &  4$\pm 1$ & This
work \\

\enddata
\end{deluxetable}

\clearpage

As for UGC 5101 and NGC 7469, the limits on the compact NIR and MIR
12$\mu$m core is found to be of similar scale --- $\la$100pc --- to the
VLBI structures.

	\vskip 0.25cm\noindent
We focus on the Southern Lobe/Interaction Region in order to estimate the
age of the radio source and, by implication, that of the QSO core.  We have
used standard synchrotron theory, assuming equipartition between
particle and field energy (minimum energy), to model the physical
characteristics of the south lobe, which we assume to be the interaction
region between a sub-relativistic jet and ambient gas:

\bul\hbox to 5.5cm{Total Energy:\hfil} $U_{Tot} \sim 10^{52} erg$

\bul\hbox to 5.5cm{Energy density:\hfil} $u \sim 2\times 10^6 erg\, cm^{-3}$

\bul\hbox to 5.5cm{Pressure:\hfil} $P_{rel} \sim 7 \times 10^{-7} dyn\,cm^{-2}$

\bul\hbox to 5.5cm{Synchrotron lifetime \@ 1.6GHz:\hfil} $\tau_{sync} \sim 2000 yr$

\bul\hbox to 5.5cm{Equipartition field:\hfil} $B_{eq} \sim 5mG$	

Assuming that the relativistic pressure in the South Lobe is balanced by 
ram pressure:

$$ P_{rel} \sim \rho v^2 $$

\noindent
We are not certain what ambient medium the jet is ramming into.  The lack
of 21-cm absorption or of maser amplification against the compact radio
source shows that the molecular disk does not cover the central source. 
\cite{UWC99} estimate the emission measure of the free-free absorbing
medium in front of the south lobe to be 10$^7$cm$^{-6}$pc and use 
the scale over which the spectral index varies to estimate the size of the
absorbing cloud to be about 5pc. This yields an electron density in the medium 

$$<n_e> \sim 10^3 cm^{-3}$$ 

\noindent
Comparson of our 1.67GHz map with their 1.4GHz map shows a spectral
index inversion which yields a consistent result for free-free absorption.
If this represents the density in the shock ionized interaction region,
then the column length
may be much smaller resulting in a correspondingly higher
electron density.  This density would be, however, a factor of 4 higher than
the density of the ambient medium.  If we take as an upper limit
$n \la 10^4 cm^{-3}$, this places a {\it lower} limit on the 
lobe-advance speed,

$$v_{adv} \ga 2 \times 10^{-4} c$$

\noindent
and thus an upper limit on the age of the recent outburst,

$$ \tau_{rs} \la 5 \times 10^5 yr $$

Considering the uncertainties in the above analysis, a comfortable age limit
might be of the order of a million years.  We have speculated
\citep{SLLD99} that this limit, combined with the suggested youth of the 
CSOs, implies that Mrk 231 is a young QSO, emerging from the shroud of dust
along the lines of the \cite{Sand88} scenario.  Of course, we do not know
that this is the first AGN-related outburst.  If the kpc-scale emission is,
indeed, produced by a jet from the nucleus, then this suggests an earlier
AGN event has occurred, probably early in the same $\sim$10$^8$ year merger
episode of which we are now witnessing the last stages.

If the advance speed in Mrk231 is as large as the speeds measured by 
\cite{T00}, then the CSO in Mrk 231 has an age of the order of only  1000 yr.
In this case the southern lobe movement since our original observations
would be approaching a milliarcsecond and will be directly measurable in
the near future.

\cite{T99} report detection of four candidate luminous radio supernovae
between 0.2--0.5mJy in Mrk 231 within a 100 pc region of the central source.
There is a faint unresolved source with flux density 0.5mJy near the
position of their brightest RSN candidate ``A'' on our image taken just
over two years earlier.  Since the flux density is comparable over the two
year interval, ``A'' appears unlikely to be an RSN and more likely to be
related to the AGN core, a possibility noted by \cite{T99}.  We note,
however, that the luminous RSN in Arp 220 exhibit longer decay times than
do ``normal'' LRSN \citep{LDLS01} so that longer timelines are necessary to
establish the nature of candidate ``A'' conclusively.  We do
not see sources near their candidates B, C or D; Taylor \etal note that
these may be artifacts.  Since the bulk of the starburst-related radio
emission appears to be coming from the HI/molecular disk, we would expect
RSN to be found in that region, as is the case with NGC 7469 \citep{C01}.

	\subsubsection{
OH Maser Emission in Mrk 231}

The OH megamaser in Mrk 231 has a spectrum typical of the class ({\it e.g.}
Staveley-Smith \etal 1987), with a velocity width of several hundred km/s.
It is one of the most luminous OH megamasers known ($10^{2.9}\,L_{\odot}$),
and was one of the first to be discovered (Baan 1985).  As
described above, Mrk 231 also exhibits strong, compact continuum emission,
which we would expect to be amplified by any diffuse foreground OH masing
screen, yielding bright maser structures on VLBI scales; based on our
picture of the geometry of the inner regions of Mrk 231, we might expect
compact OH megamaser amplification spots against the northern radio
continuum lobe, but not against the core and southern lobe, to which we
apparently have a clear line of sight to the south of the inclined HI and
molecular disk.  We might also expect compact high-gain (apparently
saturated) masers of the type discovered in Arp 220 \citep{LDSL98},
IIIZw35, and IRAS 17208$-$0014 \citep{DLSL99} for which no detectable
background continuum source exists.

To our surprise, no such compact structures of either sort exist in Mrk 231, 
on scales significantly smaller than $\sim$0.1 arcseconds.  At the locations 
of the brightest continuum emission, upper limits on the maser amplification 
ratio $S_{line}/S_{cont}$ are of order 0.05.  At the location of the 
northern lobe, the lack of detectable compact maser emission gives an 
amplification ratio limit of order unity.

Our OH image of Mrk 231, integrated over the 1667MHz line, is shown in
Figure 4.  The diffuse OH emission we detect is poorly imaged due to the
paucity of short baselines in our array, but based on the information thus
gleaned, the location and extent of the emission is consistent with an
origin in the molecular disk of material identified by \cite{BS96}, and is
likely to be amplified radio continuum emission from this disk, emission
which was imaged by \cite{T99}.  The OH emission in Mrk 231, which lacks
any compact component, is thus consistent with the classical OH megamaser
model for diffuse emission originally presented by \cite{Baan85}.

\clearpage

\begin{figure}
\figurenum{4}
\plotone{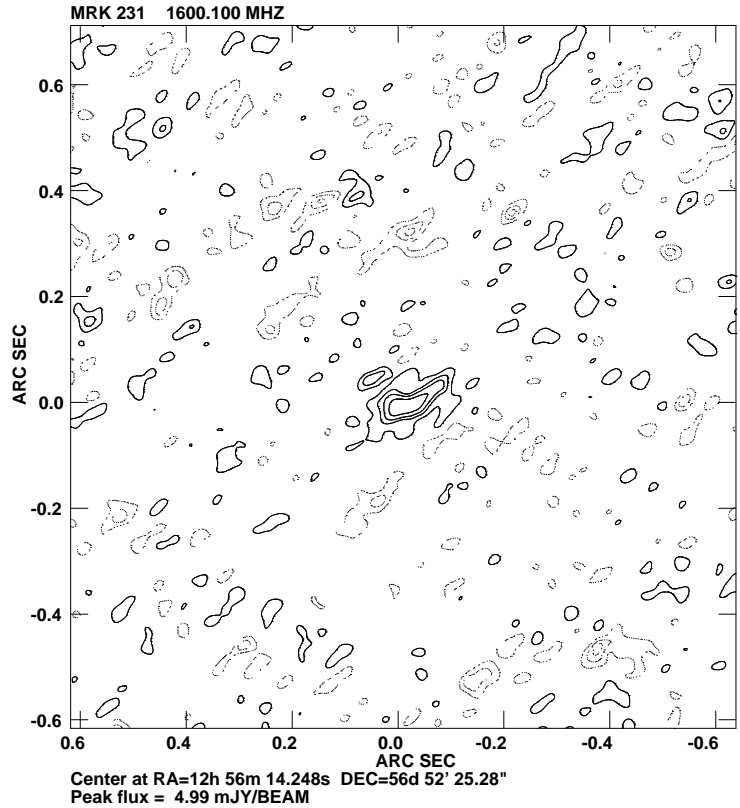}
\caption{
OH 1667MHz maser image of Mrk 231.  Axes are the RA and DEC offsets from
the continuum core center.  This image was made by averaging all
frequency channels with continuum-subtracted maser emission, discarding the
longer baselines, and tapering the (u,v) data to improve angular
resolution, thereby improving the surface brightness sensitivity relative
to a full resolution image.  Contours are -2.7, -1.9, -1.0, 1.0, 1.9,
2.7, 3.7 mJy.}

\end{figure}

\clearpage

	\section{
Discussion}

In each of the three galaxies our imaging results have demonstrated that
the radio continuum on scales of $\sim$100pc is dominated by an AGN, not a 
starburst.   Our VLBI imaging shows that the compact emission is 
characterized by well-defined ridgelines and high brightness yet spatially 
resolved components.  The structure and flux densities of these VLBI 
components are not consistent with starburst-generated radio supernovae of 
the type found in Arp 220.

In all three cases, an unresolved AGN core has also been detected in
the NIR and MIR, with upper limits comparable to the size
of the AGN-dominated 18cm core structure.  

Our direct measurements of the 18cm AGN-related core radio flux for these
three objects are 38$\pm$4 mJy for UGC 5101, 15$\pm$1 mJy for NGC 7469 and
91$\pm$9 mJy in Mrk 231.  For UGC 5101 and NGC 7469 our measured values may
be an underestimate of the total flux from the AGN due to flux on
intermediate scales missed by our long baselines.  For NGC 7469 a better
estimate of the AGN flux is the 30 mJy detected in the compact source by
MERLIN by \cite{TGPK01}, which is twice as high as the 15 mJy detected by
the longer baselines of our VLBA map.

\subsection{The Circumnuclear Regions}

We emphasize that, in addition to AGN activity, each LIG is characterized
by an intense starburst in the central regions.  However since we have
clearly demonstrated that the inner 100pc or so is dominated at 18cm by the
AGN in each case, the starburst is a circumnuclear one, and there must be
an interface region between the two phenomena.
 
To investigate the relationship of the AGN and starburst in these three
galaxies we have computed the ``infrared-to-radio parameter'' {\it q}, the
ratio of infrared to radio 20cm flux density (CHYT), as a guide to the
infrared emission expected from the starburst. {\it q} has been found to
have a remarkably limited dispersion over 5 orders of magnitude in
luminosity for star-forming regions and galaxies \citep{Y01}.  Table 5
presents the results of this calculation.

As a first estimate for {\it q} for the starburst, $q_{sb}$, we assume
that:
1) the VLBI emission, $S_{vlbi}$ (column 9, corrected by 8\% from 18cm to
20cm assuming a spectral index of 0.7), is produced by the AGN, 2) more
extended emission, $S_{vla}-S_{vlbi}$ (column 8 $-$ column 9) is produced
by the circumnuclear starburst, 3) the unresolved nuclear mid-infrared
(12$\mu$m) component is produced by the AGN, while 4) extended mid-infrared
emission is starburst related.  
Assuming that the mid-infrared component ratios reflect those in
the far-infrared we obtain the Starburst fraction (column 5) and AGN
fraction (column 6) and the Starburst Luminosity (column 3) and AGN
Luminosity (column 4).  Since we expect the AGN component to be warmer in
the infrared, this represents a lower limit to the Starburst fraction and
Luminosity.  It is worth noting that \cite{F03} obtain a very similar
Starburst fraction (60\%) for Mrk 231 based upon SED modeling.

The resulting values of $q_{sb}$ are given in column 10 of Table 5.  The
values for all three galaxies are significantly low, {\it i.e.} relatively
radio-loud, compared to a canonical average value of 2.34 (CHYT) for 
star-forming galaxies.  This may indicate that we have underestimated the
fraction of the bolometric luminosity arising in the starburst vs the AGN;
however, it would require essentially {\it all} of the measured IR
luminosity of NGC 7469 and Mrk 231 to force $q_{sb}$ to 2.34, and for UGC
5101 $q_{sb}$ would still be 50\% low: 2.17.  Such a solution is implausible
given the MIR core luminosities detected directly in each of the 3 LIGs
\citep{So00, So02, Miles96}.

One obvious way to reconcile the low $q_{sb}$ values is to assume that some
of the radio emission on larger size scales, which we have attributed to the 
starburst, in fact arises ultimately from the AGN instead, either from
unrecognized jet structure or energy pumped into the dense ISM from the
central AGN in a more spatially distributed manner which is unrecognized
in the VLA maps.   To follow this argument to a logical limit we
derived the total radio flux that must be attributed to the AGN instead
of the starburst in order to raise $q_{sb}$ to 2.34; the result is given
in column 11 of Table 5.   We find we would have to multiply the
VLBI-scale detected flux of UGC 5101 by 2.3 to 93 mJy to force 
$q_{sb} = 2.34$.   The MERLIN-detected flux of NGC 7469 would have to
be tripled to 113 mJy, while the core and lobe emission of 98 mJy of 
Mrk 231 would need to be increased by a factor of 1.4 to 140 mJy (column 11
of Table 5).  

We conclude that if we expect a nominal {\it q} value of 2.34 in
the circumnuclear starbursts in these three galaxies, then the radio flux
due to the AGN is $\sim$ 1.5 to 3 times larger than that detected on long 
baselines in the compact cores of these sources.  A similar analysis 
by Roy \etal\ (1998) and Corbett \etal\ (2002) for two different
samples of AGN with compact radio cores, has resulted in similar conclusion: 
they find evidence for radio structures associated with the compact AGN 
radio core source which extend tens to hundreds of pc around the core.   

\clearpage
 
\begin{deluxetable}{l c c c c c r r r r r}
\tabletypesize \footnotesize
\tablenum{5}
\tablewidth{0pt} 
\tablecaption{The Starburst Infrared-to-Radio Ratio, $q_{sb}^\dag$}
\tablehead{
\colhead{Galaxy} & \colhead{log L$_{IR}$} & \colhead{log L$_{SB}$} & 
	\colhead{log L$_{AGN}$} & \colhead{Starburst} & 
	\colhead{AGN} & \colhead{$q_{tot}$} & 
	\colhead{S$_{vla}$} & \colhead{S$_{vlbi}$} & 
        \colhead{$q_{sb}$} & \colhead{S$_{AGN}$} \\
\colhead{} & \colhead{(L$_\odot$)} & \colhead{(L$_\odot$)} & \colhead{(L$_\odot$)} & 
	\colhead{fraction}  & \colhead{fraction} & \colhead{} & \colhead{(mJy)} &
	\colhead{(mJy)} & \colhead{} & \colhead{(mJy)} \\
\colhead{(1)} & \colhead{(2)} & \colhead{(3)} & \colhead{(4)} & 
	\colhead{(5)}  & \colhead{(6)}  & \colhead{(7)}  & \colhead{(8)}  & 
	\colhead{(9)} & \colhead{(10)} & \colhead{(11)} \\
}
\startdata
UGC 5101 & 11.96 & 11.81 & 11.44 & 0.7 & 0.3 & 2.06 & 146 & 41.2 & 2.05 &  93 \\
NGC 7469 & 11.60 & 11.30 & 11.30 & 0.5 & 0.5 & 2.24 & 183 & 32.5 & 2.02 & 113 \\
Mrk 231  & 12.51 & 12.29 & 12.11 & 0.6 & 0.4 & 2.19 & 240 & 98.6 & 2.20 & 140 \\
\enddata

\dag {$q_{tot}$ and L$_{IR}$ are revised slightly 
from the values given by \cite{SLL98} because the IRAS fluxes of these
galaxies have been revised in a new study by \cite{Sanders03}.}

\end{deluxetable}

\clearpage

\subsection{AGN Radio Core Powers and a Comparison to Radio Quiet QSOs}

\cite{LSL95} showed that radio quiet PG QSOs follow a relationship between the
core radio power and bolometric luminosity, and that ULIRGs with VLBI-scale
emission follow a similar relation with higher scatter.  This is
illustrated in Figure 5, where we plot PG QSOs with z$<$0.6, and the new
core power estimated for the 3 LIGs, corrected to 20 cm assuming a power
law slope of 0.7, and the assumed AGN bolometric luminosities given in
Table 5.  Compared to the median radio quiet PG QSO relation (the lower
sequence of points in Figure 5; the upper sequence of points are radio-loud
PG QSOs), all three LIGs have over-luminous radio cores, by factors of
about 14, 8, and 10 for UGC 5101, NGC 7469 and Mrk 231 respectively.  If 
we plot the revised AGN-related radio fluxes resulting from the $q$
analysis above (column 11 of Table 5; small closed symbols), these LIGs will
be even more over-luminous in the radio --- these
appear to be radio-loud QSOs.

\clearpage

\begin{figure}
\figurenum{5}
\plotone{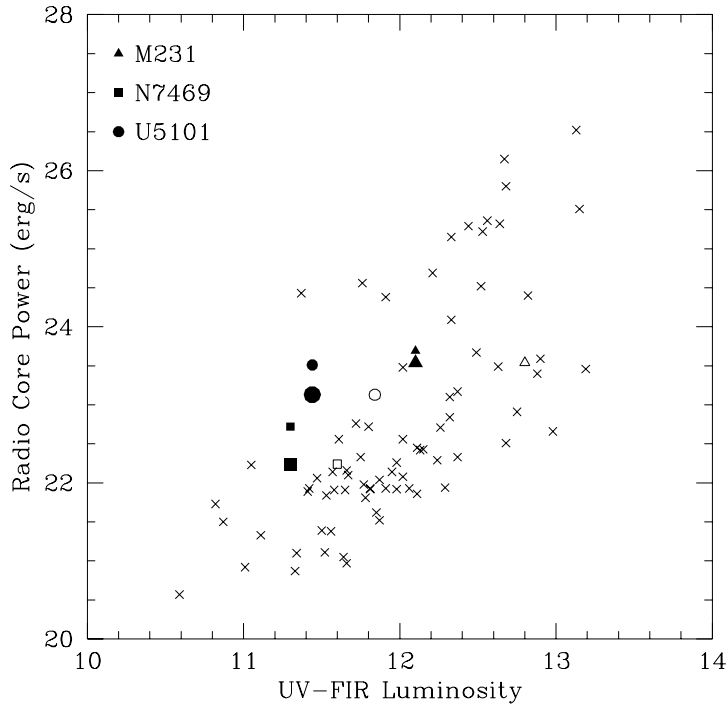}
\caption{
Integrated UV to FIR luminosity {\it vs} radio core power at 21cm 
for the three LIGs compared to PG QSOs \citep{LSL95}.  Each AGN is plotted
three times; the large filled symbols show the core radio power 
corresponding to the measured flux which we attribute to the AGN cores, 
38/15/88 mJy for UGC 5101/NGC 7469/Mrk 231, respectively, and assuming
the AGN luminosity fraction given in column 4 of Table 5.  The small 
filled symbols represent the core radio power in Column 11 of Table 5.   
The small open symbols represent the maximum
possible AGN UV-FIR luminosity in the extreme case that the contribution of 
the starburst is negligible, and that some flux has been missed altogether due
to beaming (\S 4.3).   All three AGN appear to be over-luminous
in the radio compared to PG QSOs, except possibly Mkn 231 under extreme 
assumptions.}
\end{figure}

\clearpage

The radio loudness would be reduced if the 
bolometric luminosity of the AGN has been under-estimated, however there is
not enough {\it total} measured UV-FIR luminosity (Table 5, column 2)
to shift these LIGs to the median PG QSO relation.  It
is possible some AGN luminosity has been missed
altogether due to absorption/beaming (small open symbols; see section 4.3
for analysis and discussion), which might make Mrk 231 consistent with the
PG QSO relation, but probably not the other two LIGs.
 
Are these three objects in fact radio loud AGN in the classic sense of 
high power output from the nucleus in the form of jets?  It is thought that
radio power is a function of black hole mass and accretion rate 
\citep{Lacy01, Laor00} and we certainly might expect accretion rate to be
high in these dust and gas rich objects.   Or are they 
unusually luminous in the radio because a low power AGN radio source is 
running into unusually dense ISM in the circumnuclear starburst?

\subsection{Constraining the Extinction and the Core AGN Luminosity Using 
NIR and MIR Data}

An unresolved nuclear point source has been detected in each of the three 
LIGs studied here with either HST or AO imaging in the near-IR J, H and K 
bands and also in the MIR (see Section 1 for details).  This emission
is assumed to arise in the central core regions of the AGN itself; extensive
studies of NIR and MIR SEDs of Seyfert galaxies and QSOs indicates
that the MIR emission is most likely due to a warm, dusty, molecular 
torus, and the NIR emission can arise either from the inner hot regions
of the torus from dust near 1000K, and/or from dust entrained in outflow
cones perpendicular to the accretion disk of the AGN, perhaps near the
location of the narrow line clouds \citep{ERR95, EHY95, Gran97, NIE02}.

The MIR luminosity of Seyfert galaxies and quasars is potentially a good
indicator of both the blue luminosity and the X-ray luminosity, for AGN
with modest N$_H$ columns, and presumably also the bolometric luminosity
\citep{Poll00, Haas00, Alon01}.    The near-IR luminosity is less well
correlated with the X-ray luminosity than is the MIR luminosity, probably 
due to a combination of
increased effects of extinction, but is still a useful diagnostic.
 
The analysis of the extinction to these regions is a highly complex topic.
It is well known that simple comparisons of E$_{B-V}$ from optical line ratios
to N$_H$ derived from X-ray absorption measurements often do not reflect a 
self-consistent extinction law resembling that of the Galaxy. \cite{M01}
argue that this indicates unusual dust grain compositon in 
the core regions of AGN, rather than a different line-of-sight path for 
the X-rays compared with the BLR or other explanations.   

A detailed analysis of this issue is beyond the scope of this paper.  Here we
address the specifics of the NIR and MIR emission and extinction for
our 3 LIGs, insofar as it can help to constrain the underlying AGN luminosity.

The PG QSOs investigated by \cite{LSL95} display a remarkably uniform
distribution of JHK colors.  This is illustrated in the tight and parallel 
relations between J, H and K absolute magnitudes as a function of integrated 
UV-FIR luminosity in Figure 6, where we have restricted the PG 
sample to objects with z$<$0.2 so that the NIR part of the rest frame 
spectrum is being sampled.  This behavior allows a meaningful comparison to 
the LIG colors, even though the path and extinction law of any extinction
to the LIG AGN cores may not be well understood.

\clearpage

\begin{figure}
\figurenum{6}
\plotone{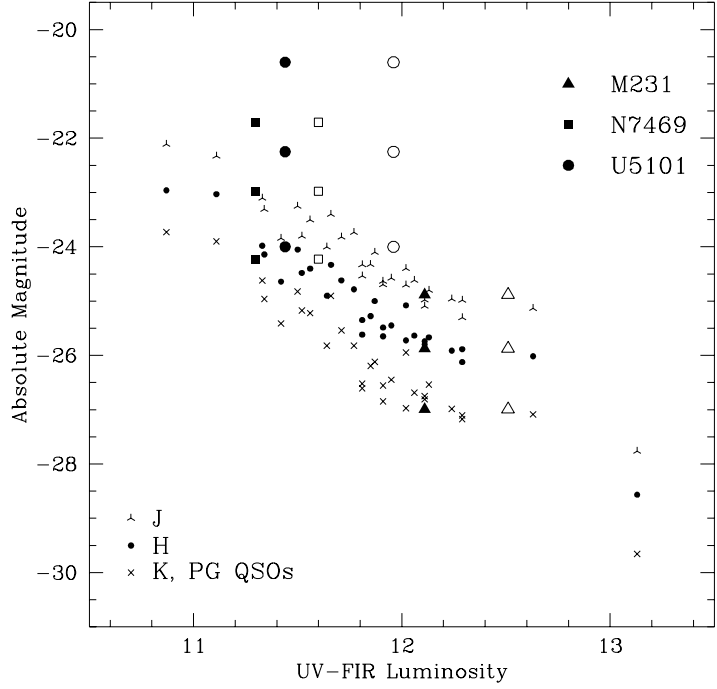}
\caption{
Absolute J, H and K core magnitudes {\it vs} integrated UV to FIR 
luminosity for the three LIGs compared to optically-selected QSOs 
(``PG QSOs'') from the Palomar-Green survey
\citep{GXN87} (small symbols, different for each band).  The 3 types of 
large symbol are different for each LIG but the same for each band; J is 
always the upper of the three symbols and K the lower.  The filled symbols 
reflect the AGN luminosity fraction in 
column 4 of Table 5 and the open symbols reflect the total UV-FIR luminosity.
The QSOs form well behaved sequences which allows an analysis of the
LIG colors.  Mrk 231's colors resemble those of the QSOs so it is probably
relatively unreddened in the NIR or has grey extinction.  Fitting to the
QSO sequence (filled triangles) produces an estimate of the QSO
UV-FIR luminosity for this galaxy, reported in column 4 of Table 5.  The
other two LIGs show evidence of extinction in their NIR colors.}
\end{figure}

\clearpage

In Figure 6 we have plotted the core J, H and K sources of each LIG twice: 
once against the total near- to far-IR luminosity of the object 
\citep{Sanders03}, and once with the AGN luminosity fraction listed 
in Table 5.   There is a quantitative difference in the location of Mrk 231 
and the other two LIGs in this figure: Mrk 231 displays the same colors as 
the PG QSOs.   This indicates that it has the same SED shape as these 
relatively unreddened QSOs, and is therefore probably itself 
essentially unreddened in the NIR.  For a Galactic extinction law, the 
reddening measured in the optical, A$_V{\sim}$2.0 mag, corresponds to
A$_K{\sim}$0.2 mag, which is unmeasurable against the dispersion of the 
PG colors in Figure 6.  The other two LIGs show larger relative separations
between the bands, and moreover [mag$_{LIG} -$mag$_{QSO}$] at the luminosity
of the LIG increases from K to H to J, in the sense expected for the effect
of reddening.   These two LIGs therefore probably display reddening in the 
NIR to the nuclear point source, that is to the core of the AGN. 
Note, however, that an analysis of the
MIR core flux in section 4.3 may amend this conclusion.

If the nuclear point source in Mrk 231 is unreddened in the NIR then it
must also suffer little extinction unless the extinction is grey.  We can
therefore use the measured NIR absolute magnitudes to fit Mrk 231
horizontally to the PG QSO relations, thereby estimating the true
luminosity of the AGN in this source, as given in Table 5: $log L = 12.11$
(L$_{\odot}$) with AGN luminosity fraction 40\%.  Consequently the
left-most points representing Mrk 231 in Figure 5 fit nicely onto the QSO
sequences because they have been forced to fit.  

For UGC 5101 and NGC 7469 we have attempted fits using a Galactic
extinction law \citep{M90}.  This analysis should only be considered
indicative of luminosities and extinctions for several reasons.  First,
the instrinsic uncertainty in the extinction due to the
dispersion in the QSO relations is large: $\sim$0.7 mag.  Also, we cannot
rule out (a) a different reddening law from Galactic; (b)
scattering which could blue the colors ({\it eg.} NGC 7469); or
(c) emission from hot dust which could redden the $[H-K]$ color, as 
suggested by \cite{Sc00} for UGC 5101.  Furthermore we may have
underestimated the bolometric luminosities of either the PG QSOs or
the LIGs by excluding X-ray emission, therefore
introducing uncertainty in the horizontal separation of the LIGs
from the QSOs.

For UGC 5101, with an assumed AGN luminosity of 11.44, we find a consistent 
fit across the three bands with A$_V$=8.9 $\pm$1.0.  For NGC 7469,
with an assumed AGN luminosity of 11.3, however, the apparent extinction 
required
to correct the J, H and K bands to the PG QSO sequence is A$_V$ = 3.2, 3.9 and
5.9 respectively, which displays inconsistency in the extinction law. 
The discrepancy from the Galactic extinction law is in the sense of the
de-reddened LIG colors being too blue.   If we take a smaller value for the 
AGN luminosity the extinction correction decreases
in size and shifts to better agreement with the Galactic law; an optimum
value occurs at about L$_{AGN}$=11.2 and A$_V$=1.9.   

\clearpage

\begin{figure}
\figurenum{7}
\plotone{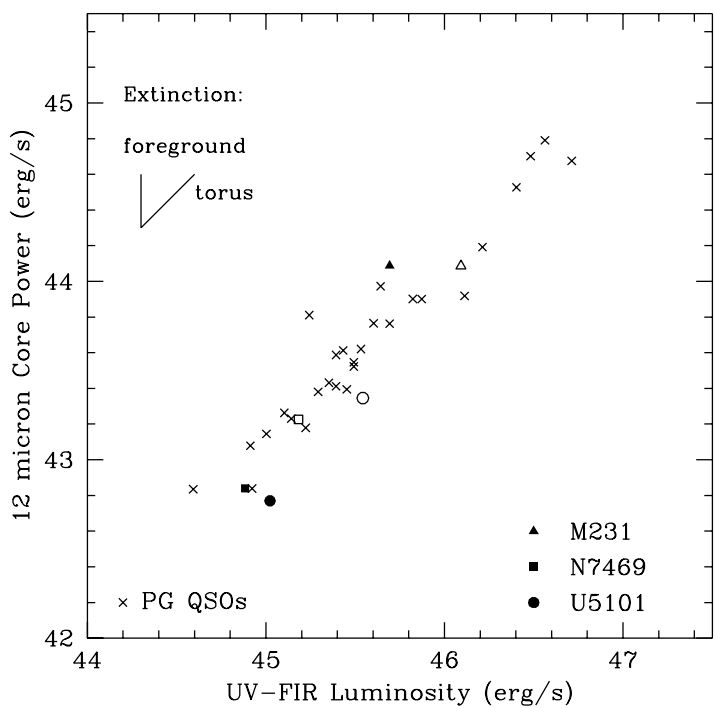}
\caption{
12 $\mu$m core power for the three LIGs  {\it vs} integrated UV to FIR 
luminosity compared to PG QSOs \citep{GXN87}.
Filled symbols reflect the AGN luminosity fraction in Table 5 column 4 while
open symbols reflect the total observed UV-FIR luminosity.  The approximate
effect of a factor of two extinction is indicated in the upper left-hand 
corner, depending upon dust geometry.  UGC 5101 appears somewhat
under-luminous compared to radio-quiet QSOs and may suffer significant 
extinction.  Mrk 231 appears slightly over-luminous.}

\end{figure}

\clearpage

Finally in Figure 7 we show the relation between integrated UV-FIR luminosity
\citep{LSL95} and core 12$\mu$m power.  For the PG QSOs we take the total IRAS
flux as reported by \cite{Sand89}, limiting the sample to z$<$0.4 so that
the 12$\mu$m filter does not shift shortward of the 7$\mu$m MIR bump.  
For the LIGs we plot the $<$100pc-scale 12$\mu$m core source (solid symbols)
\citep{So00, So02, Miles96}, taking as the AGN power the luminosity given in 
Table 5 column 4.  Also shown (open symbols) are total IRAS 12$\mu$m powers 
vs total luminosity (column 2 of Table 5). The PG QSO 
power could be overestimated if there is substantial star formation within 
the large IRAS beam \citep{Haas01}.  The PG QSOs show a very tight relation 
in this figure, with 12$\mu$m power $\sim$1\% of total power 
(neglecting the X-ray range).    

The effects of a factor of 2 extinction in Figure 7 are shown.   Since the 
source of the MIR emission is thought to be a dusty torus which can 
become optically thick in the MIR,
a detailed radiative transfer model is required to describe the source.  
Silicate absorption features are indicative of overall extinction, however
the equivalent width of such features depends not only on the intrinsic
equatorial optical depth of the torus, but on the inclination to the
line of sight, the torus height, and the dust temperature.  Inclination
is the biggest overall factor in MIR extinction, with edge-on disks
being strongly attenuated in observed luminosity.  The absorbed MIR
luminosity of such an inclined torus is not necessarily detected at
longer IR wavlengths after absorption and re-emission from more distant, 
cooler dust grains, because the torus does not emit isotropically.

In Figure 7 we see that the uncorrected 12$\mu$m core power of UGC 5101 is 
low by a factor of 2--2.5 compared to the PG QSOs (though the QSO relation is 
not strongly defined in this luminosity range) and that UGC 5101 is more
under-luminous compared to the QSOs than the other two LIGs.  This could 
either mean that there is considerable extinction at 12$\mu$m or that the
UV-FIR AGN luminosity has been over-estimated.   We favor the first 
interpretation because in large 
aperture ISO spectra, UGC 5101 shows the deepest water ice 
absorption at 6.0$\mu$m among the SEDs reported by \cite{Spoon02}, coupled 
with a very strong silicate absorption and PAH emission.   Also, the core JHK 
colors are significantly redder than for NGC 7469 and Mrk 231, as shown in
Figure 6.   Thus a consistent picture arises in which the AGN core of 
UGC 5101 is more 
heavily absorbed throughout the NIR and MIR than the other 2 LIGs, and 
in which the 12$\mu$m core luminosity in Figure 7 (filled circle) is an 
underestimate of the AGN MIR luminosity.  

Although the 12$\mu$m AGN luminosity is underestimated in Figure 7, the
{\it total} luminosity of the UGC 5101 AGN (horizontal axis) is not
necessarily affected since it is an independently inferred parameter.  
However the bolometric luminosity of the entire galaxy may
have been underestimated due to MIR extinction, depending on where the
extinction occurs, and this would affect the analysis of Section 4.1 and
the inferred total AGN luminosity.  Specifically, if the extinction occurs
in front of the torus in the starburst medium, the absorbed MIR flux will
have been accounted for in the total far-infrared emission detected by IRAS
(open circle in Figure 7) and the Section 4.1 analysis is unaffected.
However, if the observed MIR extinction occurs within a torus inclined
partly into our line-of-sight, it is possible the some of the missed flux
is beamed out of our line-of-sight altogether.  In that case the total
galaxy luminosity is underestimated by IRAS (column 2 of Table 5), the
starburst vs AGN luminosity fraction would decrease in the analysis of
Section 4.1 (columns 5 and 6), and the total AGN luminosity of column 4
would increase.  Thus both data points in Figure 7 would shift up and to
the right.

NGC 7469 is more consistent with the QSO relation in Figure 7, though perhaps 
a little under-luminous.     
\cite{Alon01} have modelled the NIR and MIR emission of NGC 7469 with
the torus models of \cite{ERR95}, finding that the flat SED requires a 
viewing angle $<$30$^{\circ}$, {\it ie.} close to face on.  In
that case little of the MIR emission would be lost to extinction (see 
Figure 9 of \cite{Alon01}).   This is generally consistent with the small,
but detectable, NIR extinction inferred from the nuclear NIR colors 
(Figure 6).

In contrast, for Mrk 231 the core 12$\mu$m power is somewhat over-luminous 
unless we have underestimated the bolometric AGN luminosity by a factor of 
about 2.5.   It's unlikely that this missing luminosity is represented by
the total UV-FIR measurement (open triangle) because that would leave nothing
over for the starburst, therefore this missing luminosity would need to be 
attributed to power missed by the IRAS measurements, {\it eg.} high 
frequency UV-X-ray flux beamed away from the line-of-sight.  There is indeed
evidence for a very high X-ray absorbing column to the core of Mrk 231 
\citep{Gal02}, which is probably sufficient to hide a nuclear
hard X-ray source of $\ga$ 10$^{44}$ erg$\,$s$^{-1}$, as implied by the 
MIR power by Figure 17 of \cite{Alon01}.

Moreover, the MIR SED of Mrk 231 is probably significantly absorbed,
exacerbating the apparent underestimation of the core AGN power deduced
from Figure 7.  Mrk 231 has a very similar 2--20$\mu$m SED shape to NGC
7469 \citep{Spoon02, So00}, except for the significant silicate absorption
feature.   To reproduce such a feature there could be foreground extinction
by the starburst medium, or the torus could be significantly more inclined 
to the line-of-sight, or of higher equatorial optical depth, 
in which case the MIR luminosity of the torus that we observe may 
underestimate the face-on luminosity by a factor of 2 or 
significantly more.     Such a result would seem at face value to be 
inconsistent with the conclusion from Figure 6 that the NIR core source is 
relatively unreddened.   However such a flat 1$<{\lambda}<$7$\mu$m SED 
\citep{So00} can instead be explained as additional warm dust emission from 
the AGN cone, as modelled for NGC 1068 \citep{EHY95}.  

In summary, there is evidence from the NIR and MIR observations that the
AGN power of Mrk 231 may be underestimated in Table 5 (column 4) by a
(highly uncertain) factor of about 5. Similarly the AGN power for UGC 5101
may be underestimated by a factor of two and overestimated by a similar
factor for NGC 7469 though a factor of two is well within the uncertainty
of this analysis.  This AGN power may be unaccounted for in the total
galaxy luminosities listed in column 2 also; this is almost certainly the
case for Mrk 231. This unaccounted AGN power in Mrk 231 may be enough to
bring it into line with the most luminous radio quiet QSOs in Figure 5
(open symbols), but UGC 5101 and NGC 7469 remain radio-loud in that plot.
Detailed hard X-ray spectroscopy and modelling will be required to resolve
this question.

\subsection{The OH Megamaser}

Among OH megamaser sources which have been studied at milliarcsecond 
resolutions, Mrk 231 is currently unique in the lack of compact structures.
Well-studied sources such as Arp 220 \citep{LDSL98} and IIIZw35 
\citep{DLSL99} have compact maser structures which account for many tens of
percent of the total OH luminosity.  Darling and Giovanelli (2002a) report
temporal variability in one distant OH megamaser, and cite further evidence
for widespread saturated (and thus compact) maser
emission in $z > 0.1$ megamaser galaxies \citep{DG02b}.  

Clues to this unique lack of compact OH megamaser emission in Mrk 231 may
come from understanding other ways in which Mrk 231 is unusual. The most 
obvious one is that the AGN core is relatively unobscured in the optical-NIR,
suggesting that it lies further along the presumed starburst-AGN evolutionary
track.    We speculate that as the AGN injected copious quantities of both 
mechanical and radiative energy into the original dense circumnuclear shroud,
as evidenced by the BAL wind and the radio jets, much of the cool molecular
environment most conducive to the formation of compact masers was destroyed.
This destruction is likely to have been complete in the AGN outflow cones.  
Mrk 231 may be a rare transition object in which the compact masers have 
been destroyed, but some diffuse maser clouds in the outer disk have so far 
survived.

However a strong circumnuclear starburst remains, which is extemely
luminous in its own right and which might be expected to contain compact
masers like those found in the Arp 220 starburst region.  This region,
roughly perpendicular to the AGN outflow axis, and of $\sim$ 400 pc
dimension, has not been cleared out by the AGN; indeed a dense optically
thick molecular torus is thought to still exist, along with a high column,
low ionization BAL wind at smaller radii.  Thus the lack of detectable OH
emission coincident with the northern continuum lobe is particularly
surprising, since the starburst disk geometry proposed by other workers
suggests that part of the disk should lie in front of that lobe, which
exhibits strong free-free absorption. 

One factor which might influence the existence of compact maser clouds is
the overall density in the Mrk 231 starburst.  Unlike Arp 220, it extends
about $\sim$400pc in diameter, and thus has a star-formation surface
density of $2 \times 10^{-4}$M$_\odot$yr$^{-1}$pc$^{-2}$, a factor of
approximately 16 less than Arp 220.  These lower density conditions in the
later stages of a starburst may inhibit the formation or pumping of compact
maser clouds.  Additionally, the lack of amplified continuum VLBI
structures may be explained by the liklihood that the cooler, outer regions
of the disk will harbor most of the OH maser clouds.  These outer regions
may lie, in projection, to the north of all the continuum VLBI-scale
structure.

	\begin{acknowledgements}
Thanks to Chris Carilli for providing us with a digital 1.4GHz image of
Mrk 231 for comparison with our data and to Tom Soifer and Jamie Bock for
discussing their mid-infrared imaging of NGC 7469. CJL and HES thank the
NRAO VLA AOC for hospitality and assistance.  HES wishes to express gratitude
to IPAC for providing continued support.  IPAC/JPL is supported by NASA.
Haystack is supported by the NSF via NEROC.  Support of this project at
UCSD has been provided by the NSF.  The National Radio Astronomy
Observatory is a facility of the NSF operated under cooperative agreement
by Associated Universities, Inc.  This research has benefitted from the
NASA/IPAC Extragalactic Database, which is operated by JPL/Caltech under
contract with NASA.
\end{acknowledgements}

\end{document}